\documentclass{article}
\usepackage[utf8]{inputenc}
\usepackage{amsmath}
\usepackage{amsfonts}
\usepackage{amssymb}
\usepackage{amsthm}

\title{Issues of Lorentz-invariance in $f(T)$ gravity\\ and calculations for spherically symmetric solutions}
\author{Alexey Golovnev\\ \\
\it Centre for Theoretical Physics, The British University in Egypt\\ \it 11837 El Sherouk City, Cairo Governorate, Egypt\\  \small agolovnev@yandex.ru}
\date{}

\begin{document}

\maketitle

\abstract{The $f(T)$ gravity is nowadays being widely used for cosmological model building, as well as for constructing spherically symmetric solutions. In its classical pure tetrad formulation it violates the local Lorentz symmetry in the space of tetrads. By using an appropriate spin connection it can be brought to a formally Lorentz invariant shape. However, despite some mathematical elegance obtained and new options of looking for non-standard couplings to matter probably made possible, it is fully equivalent in itself to the initial formulation. It seems that this fact, that the covariantisation does not in principle change anything, is not understood well by the community. Therefore we give a pedagogical introduction to these topics. And, on top of that, we show that obtaining the explicit equations for spherically symmetric solutions in $f(T)$ gravity is not very difficult computationally and can be easily done even without any computer at hand.}

\section{Introduction}

As is well-known by now, the gravitational interaction can be equivalently described by torsion or non-metricity instead of the curvature \cite{BHK}. Of course, the fact that the matter apparently tends to move along the Levi-Civita geodesics does not give much support for using torsion or non-metricity as fundamental variables of gravity without curvature, and therefore these equivalent descriptions are definitely not natural enough to simply substitute the usual general relativity. However, they can naturally provide us with modifications of gravity very different from those which are usually obtained from its standard formulation. And since in the modern theoretical physics we have many perfectly good reasons for trying to modify the theory of gravity, we need a better understanding of teleparallel approaches to the theory of this interaction.

The teleparallel description of gravity starts from taking a tetrad of vectors $e^{\mu}_a$ whose inverse (as a matrix) $e^a_{\mu}$ gives the metric as $g_{\mu\nu}=\eta_{ab}e^a_{\mu}e^b_{\nu}$. However, it then goes in terms of the flat connection
\begin{equation}
\label{simpcon}
\Gamma^{\alpha}_{\mu\nu}=e_a^{\alpha}\partial_{\mu}e^a_{\nu}
\end{equation}
which obviously breaks the local Lorentz invariance. In the action of TEGR, the teleparallel equivalent of general relativity, this non-invariance resides in merely a surface term, while the rest is just the same as the Einstein-Hilbert action. Therefore, no difference at the level of classical equations of motion. 

However, one can always go for generalisations. In particular, one of the most popular ones is known under the name of $f(T)$ gravity \cite{FeFi}. As is obvious from the name, it takes a non-linear function of the Lagrangian density of TEGR. Of course, the former boundary term is no longer a boundary term then, and it affects the equations of motion which acquire an antisymmetric part. Therefore, $f(T)$ is a model of modified gravity which, in its pure tetrad (no spin connection) formulation, breaks the local Lorentz invariance in the space of tetrads.

As always, a kind of St{\" u}ckelberg trick can be used to restore the invariance \cite{Martin, GKS}. This covariantisation simply rewrites the theory in a different way without changing the physical properties. It gives more theoretical elegancy and might help with calculations sometimes, but it does not solve any particular problem. The equation for the spin connection is simply the antisymmetric part of the equation for the tetrad, and the spin connection itself can always be made equal to zero by a gauge choice. This is fully generic for any modified teleparallel model covariantised by this method, not only for $f(T)$.

Unfortunately, it seems that the research community has not yet reached a common opinion on the properties of the covariant approach to teleparallel gravity. The viewpoints range from taking the covariant approach as the only meaningful one \cite{spinc}, and all the way to expressing certain worries that it might even be problematic \cite{Argent}. In some recent papers on spherically symmetric solutions one can notice the statements that the covariant approach is important because it allows to avoid the problem of non-uniqueness of the tetrad \cite{SariCai, RNT}. This is however totally wrong. And this is the topic of the present paper.

Let us formulate the claims more accurately. In TEGR, one can always use any possible tetrad for a given metric, and all of them would be equivalent in the sense of producing one and the same solution. This local Lorentz symmetry is broken in the $f(T)$ models. The problem is that it can still allow for a number of different tetrads with the same metric, and those would generically give physically different solutions. This is not very surprising since the tetrad becomes a more fundamental variable than the metric, and it has more information. The worry of many people was that this preferred frame effect is annoying, and the covariantisation aims at providing us with ability to use any tetrad we like for a given metric. All these tetrads are then equivalent to each other if, when going from one to another, we also make the corresponding transformation of the spin connection. However the preferred frame effect is still there, in a different disguise now. It is just that the model can allow for different spin connections with one and the same tetrad, which again means that physically different solution with the same metric can be obtained.

In Section 2 I briefly present the basics of this type of gravity. In Section 3 I review what is currently known about its spherically symmetric solutions, and also I present all the details of calculations which lead to that. In Section 4 I explain what diffeomorphism invariance means for these issues. In Section 5 I describe the role of local Lorentz invariance or lack of it when comparing different formulations of $f(T)$ gravity. Finally, in Section 6 I conclude.

\section{The f(T) gravity}

The quest for TEGR action can start from observing that a metric-compatible connection $\Gamma^{\alpha}_{\mu\nu}$ with torsion differs from the Levi-Civita one $\mathop\Gamma\limits^{(0)}{\vphantom{\Gamma}}^{\alpha}_{\mu\nu}$ by a contortion tensor:
\begin{equation*}
\Gamma^{\alpha}_{\mu\nu}=\mathop\Gamma\limits^{(0)}{\vphantom{\Gamma}}^{\alpha}_{\mu\nu}(g)+K^{\alpha}_{\hphantom{\alpha}\mu\nu}
\end{equation*}
which is defined in terms of the torsion tensor $T^{\alpha}_{\hphantom{\alpha}\mu\nu}=\Gamma^{\alpha}_{\mu\nu}-\Gamma^{\alpha}_{\nu\mu}$ as
\begin{equation}
\label{cont}
K_{\alpha\mu\nu}=\frac12\left(T_{\alpha\mu\nu}+T_{\nu\alpha\mu}+T_{\mu\alpha\nu}\right).
\end{equation}
It is antisymmetric in the lateral indices because I ascribe the left lower index of a connection coefficient to the derivative, e.g. $\bigtriangledown_{\mu}T^{\nu}\equiv\partial_{\mu}T^{\nu}+\Gamma^{\nu}_{\mu\alpha}T^{\alpha}$.

The curvature tensor
\begin{equation}
\label{curvature}
R^{\alpha}_{\phantom{\mu}\beta\mu\nu}=\partial_{\mu}\Gamma^{\alpha}_{\nu\beta}-\partial_{\nu}\Gamma^{\alpha}_{\mu\beta}
+\Gamma^{\alpha}_{\mu\rho}\Gamma^{\rho}_{\nu\beta}
-\Gamma^{\alpha}_{\nu\rho}\Gamma^{\rho}_{\mu\beta}
\end{equation}
for the two different connections obviously has a quadratic in $K$ expression in the difference. Then making necessary contractions, such as $R_{\mu\nu}=R^{\alpha}_{\phantom{\alpha}\mu\alpha\nu}$ and $R=g^{\mu\nu}R_{\mu\nu}$, we can come to the basic result \cite{MeRev}
\begin{equation}
\label{relation}
\mathop{R}\limits^{(0)}+{\mathbb T}+2\mathop{\bigtriangledown_{\mu}}\limits^{(0)}T^{\mu}=0
\end{equation}
since the connection $\Gamma$ has zero curvature. Here $T_{\mu}\equiv T^{\alpha}_{\hphantom{\alpha}\mu\alpha}$ is the torsion vector while the torsion scalar
$${\mathbb T}\equiv\frac 12 S_{\alpha\mu\nu}T^{\alpha\mu\nu}$$
is given in terms of the superpotential
\begin{equation}
\label{superpot}
S_{\alpha\mu\nu}\equiv K_{\mu\alpha\nu}+g_{\alpha\mu}T_{\nu}-g_{\alpha\nu}T_{\mu}
\end{equation}
which is antisymmetric in the last two indices, the same as the torsion tensor.

Due to the relation (\ref{relation}), the Einstein-Hilbert action $-\int d^4 x \sqrt{-g}\mathop{R}\limits^{(0)}$ is equivalent to the TEGR one, $\int d^4 x \| e\| \mathbb T$. They are the same, up to the surface term ${\mathbb B}\equiv 2\mathop{\bigtriangledown_{\mu}}\limits^{(0)}T^{\mu}$. Of course, this equivalence disappears when we go to modified gravity, for example the $f(T)$ gravity:
$$S=\int f(\mathbb T)\cdot \| e\| d^4 x.$$

\subsection{A geometric way to equations of motion}

Again thanks to the basic relation (\ref{relation}), the variation of the $f(T)$ action can be in a big part reduced to the usual Riemannian one \cite{GK}:
$$\delta S=\int\left(f\delta \| e\|-\| e\|f_{T}\left(\delta\mathop{R}\limits^{(0)}+2\delta\left(\frac{1}{\| e\|}\partial_{\mu}(\| e\|T^{\mu})\right)\right) \right)$$
with $\delta \| e\|=\| e\|e^{\mu}_{a}\delta e^a_{\mu}$. We presented the structure of the Levi-Civita divergence explicitly in order to stress that its notion must also be subject to variation. Note that it has been missed in our initial papers on this topic \cite{GKS, MeRev} leading to a mistake in the equation of motion. 

Now the only unusual piece of work, for anyone who has ever worked with general relativity, is variation of the torsion vector. A convenient way, also for many other modified teleparallel models, would be to look at the following formula:
$$\bigtriangledown_{\mu}(e^{\alpha}_a\delta e^a_{\nu})=\partial_{\mu}(e^{\alpha}_a\delta e^a_{\nu})+\Gamma^{\alpha}_{\mu\rho} e^{\rho}_a \delta e^a_{\nu}-\Gamma^{\rho}_{\mu\nu} e^{\alpha}_a \delta e^a_{\rho}$$
where we can see that $\delta e^a_{\nu}\partial_{\mu} e^{\alpha}_a=-(\delta e^a_{\nu}) e^{\alpha}_b e^{\rho}_a\partial_{\mu} e^b_{\rho}=-\Gamma^{\alpha}_{\mu\rho} e^{\rho}_a \delta e^a_{\nu}$ and also that $\Gamma^{\rho}_{\mu\nu} e^{\alpha}_a \delta e^a_{\rho}=e^{\rho}_b(\partial_{\mu}e^b_{\nu}) e^{\alpha}_a \delta e^a_{\rho}=-(\partial_{\mu}e^b_{\nu})\delta e^{\rho}_b$ which finally gives
$$\bigtriangledown_{\mu}(e^{\alpha}_a\delta e^a_{\nu})=(\delta e^{\alpha}_a)\partial_{\mu}e^a_{\nu}+e^{\alpha}_a\partial_{\mu}(\delta e^a_{\nu}).$$
Since $\Gamma^{\alpha}_{\mu\nu}=e^{\alpha}_a\partial_{\mu}e^a_{\nu}$, we obtain a very nice formula for the variation of the flat connection coefficient \cite{GK}:
\begin{equation*}
\delta \Gamma^{\alpha}_{\mu\nu}=\bigtriangledown_{\mu}\left(e^{\alpha}_a \delta e^a_{\nu}\right)=\mathop{\bigtriangledown_{\mu}}\limits^{(0)}\left(e^{\alpha}_a \delta e^a_{\nu}\right)+K^{\alpha}_{\hphantom{\alpha}\mu\beta}e^{\beta}_a \delta e^{a}_{\nu}-K^{\beta}_{\hphantom{\beta}\mu\nu}e^{\alpha}_a \delta e^{a}_{\beta}.
\end{equation*}
It can be used in any modified teleparallel model, but in the case of $f(T)$, or even $f(T,B)$, we just need to specify it for the variation of the torsion vector \cite{GKS, GK}:
\begin{equation*}
\delta T_{\mu}=\partial_{\mu}\left(e^{\nu}_a\delta e^{a}_{\nu}\right)-\mathop{\bigtriangledown_{\alpha}}\limits^{(0)}\left(e^{\alpha}_a \delta e^a_{\mu}\right)-K^{\alpha}_{\hphantom{\alpha}\alpha\nu}e^{\nu}_a \delta e^{a}_{\mu}+K^{\nu}_{\hphantom{\nu}\alpha\mu}e^{\alpha}_a \delta e^{a}_{\nu}.
\end{equation*}

After some little exercise, and cancelling an overall factor of $e^{\mu}_a$, the equation of motion can be written in vacuum as \cite{GK, oldgood}
\begin{equation}
\label{eom}
f_{T}\mathop{G_{\mu\nu}}\limits^{(0)}+\frac12 \left(f-f_{T}{\mathbb T}\right)g_{\mu\nu}+f_{TT}S_{\mu\nu\alpha}\partial^{\alpha}{\mathbb T}=0.
\end{equation}
The higher derivatives from the variation of the scalar curvature got cancelled by those from the variation of divergence of the torsion vector, as it must be due to the lower derivative order of the $f(T)$ action. 

Note that in the case of linear function $f$ we get the usual general relativity with renormalised (by the factor of $f_T$) gravitational constant and a cosmological constant given by $f(0)$. Otherwise $f_{TT} \neq 0$ and the equation also has antisymmetric part. This is because the local Lorentz transformations of the tetrad do have some non-trivial effect on the action. However, in case of constant ${\mathbb T}$ the solutions obviously reduce to the general relativistic ones for any function $f$.

In presence of matter coupled to the metric in the usual way, we would need to put the energy-momentum tensor to the right hand side of the equation (\ref{eom}). Of course, a certain problem for teleparallel theories is to have fermions. It can be safely done by coupling them to the spin connection which corresponds to the Levi-Civita spacetime connection. It's fine, but kind of inelegant for this approach.

\subsection{Diffeomorphism invariance}

This model of gravity is an unusual one. Taken without a spin connection, what is called pure tetrad formulation, it breaks the local Lorentz invariance in the space of tetrads. However, we have to bare in mind that it is invariant under diffeomorphisms.

Indeed, all the quantities which could have raised any doubt are constructed from the torsion tensor. But looking at the definition
$$T^{\alpha}_{\hphantom{\alpha}\mu\nu}=e_a^{\alpha}(\partial_{\mu}e^a_{\nu}-\partial_{\nu}e^a_{\mu})$$
and remembering that under a change of coordinates every $e^a_{\mu}$ transforms as a 1-form, while $e_a^{\alpha}$ -- as a vector, we see that it is indeed a genuine tensor, from the viewpoint of diffeomorphisms.

As will be discussed below, in Section 4, it is a very important property of the model, not only from the fundamental ideas that coordinates are not allowed to play any foundational physical role, but also even for seeing that some interesting solutions are indeed possible. On the other hand, given that a tetrad is a set of four vectors, it being diagonal or not diagonal as a matrix also depends on the chosen coordinates. Therefore, the usual claims of impossibility to use a diagonal tetrad for spherically symmetric solutions in $f(T)$ gravity also depend on the implicit assumption that the spherical coordinates are used.

\subsection{Covariantisation with respect to Lorentz}

As we have mentioned above, the local Lorentz invariance is broken in models of $f(T)$ gravity. This is because a Lorentz matrix acting on the tetrad would itself be acted upon by the derivative in the definition of the teleparallel connection (\ref{simpcon}). One can correct it by introducing the spin connection $\omega$:
\begin{equation}
\label{invcon}
\Gamma^{\alpha}_{\mu\nu}=e_a^{\alpha}\left(\partial_{\mu}e^a_{\nu}+\omega^a_{\hphantom{a}\mu b} e^b_{\nu}\right)    
\end{equation}
where it is assumed to be flat
\begin{equation*}
\partial_{\mu}\omega^a_{\hphantom{a}\nu b}-\partial_{\nu}\omega^a_{\hphantom{a}\mu b}+\omega^a_{\hphantom{a}\mu c}\omega^c_{\hphantom{c}\nu b}-\omega^a_{\hphantom{a}\nu c}\omega^c_{\hphantom{c}\mu b}=0
\end{equation*}
and take values in the Lie algebra of the Lorentz group which means that $\omega_{a\mu b}=-\omega_{b\mu a}$ for $\omega_{a\mu b}\equiv \eta_{ac}\omega^c_{\hphantom{c}\mu b}$. The spacetime connection $\Gamma$ then also remains flat and metric-compatible.

Obviously, this definition of the connection coefficients appears to be invariant under the simultaneous transformation of $e$ and $\omega$ by a local Lorentz matrix $\Lambda(x)$:
\begin{equation}
\label{gauge}
e^a_{\mu}\rightarrow \Lambda^a_b e^b_{\mu}, \qquad \omega^a_{\hphantom{a}\mu b}\rightarrow \Lambda^a_c \omega^c_{\hphantom{c}\mu d} {\Lambda^{-1}}^d_b-(\partial_{\mu} \Lambda^{a}_{c}) {\Lambda^{-1}}^{c}_{b}.
\end{equation}
Therefore, all the quantities in the definition of the theory appear to be invariant under this transformation which is a gauge freedom then, in addition to diffeomorphisms.

One can start from a pure tetrad solution with $\omega=0$, and then perform all desired local Lorentz transformations which would allow for any possible tetrad for a given metric, at the price of getting all possible spin connections of the $-(\partial\Lambda)\Lambda^{-1}$ shape. On the other hand, starting from any flat Lorentzian spin connection, one can make a transformation for taking $\omega=0$ as a gauge choice, at least locally.

Globally, there can be cohomological obstructions for taking the pure tetrad gauge. And we will not discuss these issues. The very formulation of teleparallel gravity as an equivalent of GR is prone to that, since we need that the spacetime manifold is absolutely parallelisable. Let us stick to local viewpoints without much thought on topology.

The variation with respect to the spin connection has to be done carefully. It is not arbitrary, it must always remain a teleparallel connection. If we don't add any new terms to the action, then even the TEGR would get to be trivial if we require vanishing of the variation of the action with absolutely arbitrary variation of the spin connection \cite{GKS}. The variation must be in the flat class. It can be done by carefully looking at what is required for making only such variations of the action vanish, or just by substituting $\omega^a_{\hphantom{a}\mu b}=-(\partial_{\mu} \Lambda^{a}_{c}) {\Lambda^{-1}}^{c}_{b}$ (all possible spin connections which can be brought to zero value by a local Lorentz transformation) and making a variation with respect to $\Lambda$ in the Lorentz group.

Since the transformation (\ref{gauge}) is a gauge transformation, it means that any transformation of the spin connection alone, in the flat class, changes the action in the same way, with opposite sign, as the corresponding Lorentz rotation of the tetrad. Therefore, the equation of motion for the spin connection coincides with the antisymmetric part of the equation of motion for the tetrad. 

And the equation of motion for the tetrad is the same as before (\ref{eom}), with the only difference that all the quantities are computed from the full connection (\ref{invcon}). Indeed, the only thing we have to check is that the variation of the torsion vector is not changed more than this. It can be seen by looking at the variation of the connection coefficient due to variation of the tetrad: $\delta\Gamma^{\alpha}_{\mu\nu}=\bigtriangledown_{\mu}\left(e^{\alpha}_a \delta e^a_{\nu}\right)$. The left hand side now has two more terms from varying the tetrad and its inverse in the new part of the connection, $e_a^{\alpha}\omega^a_{\hphantom{a}\mu b} e^b_{\nu}$, while the right hand side also has two new terms because of the very same addition to the connection coefficients for the covariant derivative. Therefore one can easily check that the formula $\delta\Gamma^{\alpha}_{\mu\nu}=\bigtriangledown_{\mu}\left(e^{\alpha}_a \delta e^a_{\nu}\right)$ is valid for variation of the tetrad in both the pure tetrad and the covariant formulations. After that there is no difference in deriving the equation of motion.

Nothing physical is changed by adding the spin connection. It is a trick for having freedom of choosing any tetrad for a given metric in a model which actually depends on this choice. One can compare it with formally restoring the $U(1)$ gauge freedom in a theory of a complex scalar field without electromagnetic field. If instead of the vector potential we introduce covariant derivatives with just a gradient of a scalar, and no other new terms into the action, then formally the gauge symmetry is there, no new physical content is added, and the variation with respect to the new scalar implies the conservation of current which already has been there before \cite{GKS}.

Once again. If we add some new variables in a way to only compensate for unwanted changes in an action with a broken symmetry, then the equations of the new variables do nothing but just repeat some parts of the other equations. This is a very basic fact, and in the context of teleparallel theories it has been explicitly checked in some other cases, too \cite{Hoh1, Hoh2}. 

In (pure tetrad) $f(T)$ we can not freely choose a tetrad, and we want to have this freedom, maybe from just the belief which we are used to, that it is the basic freedom of choosing observers and frames. To have this freedom, we introduce the spin connection. It is done so that we can indeed equivalently use another choice of the tetrad, which might not even be allowed by the pure tetrad equations of motion, but the price to pay is that we add the spin connection, in a way to just compensate for all unwanted changes in the geometrical quantities such as the torsion tensor.

In principle, even without covariantisation we can still describe all observers we want if we agree that their tetrad is just a different thing, not the one which is used for the definition of the teleparallel connection. In other words, a tetrad used to define a reference frame should be absolutely independent of $e^a_{\mu}$. They can coincide, but we freely change the reference tetrad without affecting the fundamental one. With such an agreement, there seems to be no deep mathematical problem with the pure tetrad approach.

In our opinion, covariantisation is just an elegant way of rewriting the model, and it can make some calculations more convenient and can give insights into the properties of Lorentzian modes. On the other hand, the covariant version might also be preferable for people who want to find well-defined conserved quantities such as gravitational energy (which in our opinion does not make any good sense at all) because treating the teleparallel gravity as a gauge theory of translations (therefore giving hopes for formal conservation laws) naturally requires the covariant way of defining.

\section{Spherically symmetric solutions}

Recently, the $f(T)$ gravity has been developed quite a lot. We now have a good understanding of the theory of linear cosmological perturbations \cite{GK, mecosm} which can be successfully used for confrontation with observations \cite{wecairo1, wecairo2}. On the other hand, there are also many foundational problems with models of this kind \cite{GoGu1, GoGu2, Guz}, with even the number of degrees of freedom being not understood well and being dependent on the background. The reason is that the local Lorentz invariance is not broken strongly enough leaving a big room for remnant symmetries to play around \cite{remn} which produces the problem of strong coupling and non-constant rank of the algebra of Poisson brackets of constraints \cite{GoGu2, Guz}. But some other modified teleparallel models might appear to be more viable, and generic properties of these models are definitely worth studying.

Given that and the current interest to modified teleparallel gravity, it is very important to also study other solutions, not only the simplest cosmological ones. And in particular, spherically symmetric solutions have recently been considered in many papers \cite{spher1, spher2, spher3}. The crucial point about this topic is that usually such solutions are searched for in spherical coordinates, and it turns out with this choice of coordinates that a diagonal tetrad is not compatible with the pure tetrad $f(T)$ equations of motion.

For the metric
$$g_{\mu\nu}dx^{\mu}dx^{\nu}=A^2(r)dt^2-B^2(r)dr^2-r^2(d\theta^2+sin^2(\theta) d\phi^2)$$
the tetrad which respects spherical symmetry and is compatible with the antisymmetric part of equations of motion (\ref{eom}) is
\begin{equation}
\label{sphtet}
e^{a}_{\mu} = \left(
\begin{array}{cccc}
A(r) & 0 & 0 & 0\\
0 & B(r) \sin(\theta)\cos(\phi) & r \cos(\theta) \cos(\phi) & -r\sin(\theta) \sin(\phi) \\
0 & B(r) \sin(\theta) \sin(\phi) & r \cos(\theta) \sin(\phi) & r \sin(\theta) \cos(\phi) \\
0 & B(r) \cos(\theta) & -r\sin(\theta) & 0
\end{array}
\right)
\end{equation}
with two arbitrary functions of radius, $A$ and $B$.

Another option was considered in the literature \cite{spher4}, with the change of sign in all the elements proportional to $r$ (two columns on the right). Actually, it is not a different choice at all. Indeed, if we change the overall sign of the tetrad, it does not change anything, neither the metric nor the torsion. It means that the tetrad from the Ref. \cite{spher4} is equivalent to the tetrad (\ref{sphtet}) if the signs of $A$ and $B$ are reversed. Since these functions are arbitrary, in essence it is the same tetrad ansatz (not precisely true for the Ref. \cite{spher4} itself because they used exponential notation for their arbitrary functions).

The properties of these and similar constructions are still under active investigation \cite{add1, add2, add3}. But let us now discuss all that from the Lorentz-covariantised point of view. The diagonal tetrad 
\begin{equation}
\label{dtet}
e^a_{\mu}=\mathrm{diag} (A,B,r,r\sin\theta)
\end{equation}
doesn't work well without a spin connection, unless in TEGR $f_{TT}=0$. However, applying a Lorentz transformation
\begin{equation}
\label{Lor}
\Lambda = \left(
\begin{array}{cccc}
1 & 0 & 0 & 0 \\
0 &  \sin(\theta)\cos(\phi) &  \cos(\theta) \cos(\phi) & - \sin(\phi) \\
0 &  \sin(\theta) \sin(\phi) &  \cos(\theta) \sin(\phi) &  \cos(\phi) \\
0 &  \cos(\theta) & -\sin(\theta) & 0
\end{array}
\right),
\end{equation}
we can obtain the tetrad (\ref{sphtet}) which solves the equations and is usually called a "good" tetrad in somewhat strange a jargon. It means that taking this tetrad (\ref{sphtet}) we can go back to the diagonal one by applying $\Lambda^{-1}$, and it will still be the same solution as long as we do not forget to introduce the corresponding non-vanishing spin connection $\omega_{\mu}=-(\partial_{\mu}\Lambda^{-1})\Lambda=\Lambda^{-1}\partial_{\mu}\Lambda$.

The spin connection can then be found \cite{spinc, GoGu3}, with tangent indices $0,1,2,3$ and spacetime indices $t, r,\theta,\phi$, as
\begin{equation}
\label{omega}
\omega_{1 \theta 2}=-\omega_{2 \theta 1}=1,\qquad \omega_{1 \phi 3}=-\omega_{3 \phi 1}=\sin\theta,\qquad \omega_{2 \phi 3}=-\omega_{3 \phi 2}=\cos\theta,
\end{equation}
with all the other components vanishing. Indeed, the non-trivial part of the Lorentz rotation (\ref{Lor}) is a 3x3 matrix of unit determinant which can immediately be inverted in mind as
\begin{equation*}
\Lambda^{-1} = \left(
\begin{array}{cccc}
1 & 0 & 0 & 0 \\
0 &  \sin(\theta)\cos(\phi) &  \sin(\theta) \sin(\phi) &  \cos(\theta) \\
0 &  \cos(\theta) \cos(\phi) &  \cos(\theta) \sin(\phi) & - \sin(\theta) \\
0 & - \sin(\phi) & \cos(\phi) & 0
\end{array}
\right),
\end{equation*}
and then we have
$$\Lambda^{-1}\partial_{\theta}\Lambda=\left(
\begin{array}{cccc}
0 & 0 & 0 & 0 \\
0 &  0 &  -1 &  0 \\
0 &  1 & 0 & 0 \\
0 & 0 & 0 & 0
\end{array}
\right), \qquad\Lambda^{-1}\partial_{\phi}\Lambda=\left(
\begin{array}{cccc}
0 & 0 & 0 & 0 \\
0 &  0 &  0 &  -\sin\theta \\
0 &  0 & 0 & -\cos\theta \\
0 & \sin\theta & \cos\theta & 0
\end{array}
\right)$$
what corresponds to $\omega^2_{\hphantom{2}\theta 1}=1, \quad \omega^3_{\hphantom{2}\phi 1}=\sin\theta, \quad \omega^3_{\hphantom{2}\phi 2}=\cos\theta$ as is given above (\ref{omega}).

As a cross-check, one can very easily see that this spin connection is indeed flat because, with these non-zero components and dependence only on $\theta$, the only non-trivial thing to check is $R^a_{\hphantom{a}b \theta\phi}(\omega)=0$ for non-zero tangent indices.

\subsection{A full derivation by hand}

As we have already mentioned above, the covariantisation of teleparallel gravities does not make any change to the contents of the model, if taken at the level of the classical equations of motion, without worrying about the value of the action. We are not interested now in quantum gravity, nor in Black Hole information paradox, and therefore we can treat $\omega=0$ as a totally legal gauge choice.

However, let us note that using a non-zero spin connection can be very useful as a technical trick for doing calculations. We will now see that taking the ansatz as the diagonal tetrad (\ref{dtet}) and the spin connection (\ref{omega}) one can calculate the equations of motion without using a computer, and still without too much effort. This is because the tetrad is diagonal. This choice of tetrad is a very natural attempt, while this spin connection can be obtained by an idea to find it by assuming that the spin connection is naturally zero for a tetrad which is diagonal in Cartesian coordinates.

Let us start by doing some calculations with an arbitrary choice of the radial coordinate $r$, so that the metric and the tetrad are
$$g_{\mu\nu}dx^{\mu}dx^{\nu}=A^2(r)dt^2-B^2(r)dr^2-h^2(r)(d\theta^2+sin^2(\theta) d\phi^2),$$
$$e^a_{\mu}=\mathrm{diag} (A,B,h,h\sin\theta).$$
Since the spin connection (\ref{omega}) does not depend on $r$ and does not have non-zero $\omega_{arb}$ components, we can safely leave it the same as before.

\subsubsection{The Riemannian part}

The nice fact about equations of motion in the form of Eq. (\ref{eom}) is that the most difficult part of it just requires a usual general relativistic calculation, i.e. the Einstein tensor, even though many authors use a fully non-covariant form of equations. Let us explicitly show how these calculations can be better performed. In all the quantities we calculate below we will have spacetime indices. But, in order to make it look nicer, we will denote those indices by numbers (unlike how it has been done above for the spin connection) so that $0, 1, 2, 3$ now denote $t, r, \theta, \phi$ respectively.

The non-zero components of the Levi-Civita connection
$$\mathop\Gamma\limits^{(0)}{\vphantom{\Gamma}}^{\alpha}_{\mu\nu}= \frac12 g^{\alpha\beta}(\partial_{\mu}g_{\beta\nu}+\partial_{\nu}g_{\beta\mu}-\partial_{\beta}g_{\mu\nu})$$
can easily be calculated since one of the lower indices must be either $1$ (corresponding to $r$) or $2$ (corresponding to $\theta$), and the other two must coincide since the metric is diagonal. The answer is as follows:
$$\mathop\Gamma\limits^{(0)}{\vphantom{\Gamma}}^{1}_{00}=\frac{AA^{\prime}}{B^2}, \qquad \mathop\Gamma\limits^{(0)}{\vphantom{\Gamma}}^{1}_{11}=\frac{B^{\prime}}{B}, \qquad \mathop\Gamma\limits^{(0)}{\vphantom{\Gamma}}^{1}_{22}=-\frac{hh^{\prime}}{B^2},\qquad \mathop\Gamma\limits^{(0)}{\vphantom{\Gamma}}^{1}_{33}=-\frac{hh^{\prime}\sin^2\theta}{B^2}, \qquad \mathop\Gamma\limits^{(0)}{\vphantom{\Gamma}}^{2}_{33}=-\sin(\theta)\cos(\theta),$$
$$\mathop\Gamma\limits^{(0)}{\vphantom{\Gamma}}^{0}_{01}=\mathop\Gamma\limits^{(0)}{\vphantom{\Gamma}}^{0}_{10}=\frac{A^{\prime}}{A}, \qquad \mathop\Gamma\limits^{(0)}{\vphantom{\Gamma}}^{2}_{12}=\mathop\Gamma\limits^{(0)}{\vphantom{\Gamma}}^{2}_{21}=\mathop\Gamma\limits^{(0)}{\vphantom{\Gamma}}^{3}_{13}=\mathop\Gamma\limits^{(0)}{\vphantom{\Gamma}}^{3}_{31}=\frac{h^{\prime}}{h}, \qquad \mathop\Gamma\limits^{(0)}{\vphantom{\Gamma}}^{3}_{23}=\mathop\Gamma\limits^{(0)}{\vphantom{\Gamma}}^{3}_{32}=\cot\theta.$$

The calculation of the Ricci tensor also goes without any trouble and gives
$${\mathop{R}\limits^{(0)}}_{00}=\frac{AA^{\prime\prime}}{B^2}+2\frac{AA^{\prime}h^{\prime}}{hB^2}-\frac{AA^{\prime}B^{\prime}}{B^3}, \qquad {\mathop{R}\limits^{(0)}}_{11}=-\frac{A^{\prime\prime}}{A}+\frac{A^{\prime}B^{\prime}}{AB}+2\frac{h^{\prime}B^{\prime}}{hB}-2\frac{h^{\prime\prime}}{h},$$
$${\mathop{R}\limits^{(0)}}_{22}=-\frac{hh^{\prime\prime}}{B^2}-\frac{{h^{\prime}}^2}{B^2}+\frac{hh^{\prime}B^{\prime}}{B^3}-\frac{hh^{\prime}A^{\prime}}{AB^2}+1, \qquad {\mathop{R}\limits^{(0)}}_{33}={\mathop{R}\limits^{(0)}}_{22}\sin^2\theta,$$
and no off-diagonal components different from zero. One can make simple checks at this stage. Multiplying $A$ by some constant does not change the spatial components but multiplies the temporal one by the square of this constant. This is because it is just a rescaling of the time variable. Having kept arbitrary $h(r)$ also allows to check the behaviour under changes of the radial variable. This is a bit more complicated. However, it is easy to see that taking Minkowski space with an arbitrary radial coordinate as $A=1$ and $B=h^{\prime}$ we get the zero Ricci tensor as it must be.

We finalise the Riemannian part of the calculations by finding the Ricci scalar
$$\mathop{R}\limits^{(0)}=2\frac{A^{\prime\prime}}{AB^2}-2\frac{A^{\prime}B^{\prime}}{AB^3}+4\frac{h^{\prime}A^{\prime}}{hAB^2}-4\frac{h^{\prime}B^{\prime}}{hB^3}+4\frac{h^{\prime\prime}}{hB^2}+2\frac{{h^{\prime}}^2}{h^2B^2}-2\frac{1}{h^2}$$
and obtaining the Einstein tensor ${\mathop{G}\limits^{(0)}}{\vphantom{G}}^{\mu}_{\nu}={\mathop{R}\limits^{(0)}}{\vphantom{R}}^{\mu}_{\nu}-\frac12 \mathop{R}\limits^{(0)} \delta^{\mu}_{\nu}$ as
$${\mathop{G}\limits^{(0)}}{\vphantom{G}}^0_0=2\frac{h^{\prime}B^{\prime}}{hB^3}-2\frac{h^{\prime\prime}}{hB^2}-\frac{{h^{\prime}}^2}{h^2B^2}+\frac{1}{h^2}, \qquad {\mathop{G}\limits^{(0)}}{\vphantom{G}}^1_1=-2\frac{h^{\prime}A^{\prime}}{hAB^2}-\frac{{h^{\prime}}^2}{h^2B^2}+\frac{1}{h^2},$$
$${\mathop{G}\limits^{(0)}}{\vphantom{G}}^2_2={\mathop{G}\limits^{(0)}}{\vphantom{G}}^3_3=-\frac{h^{\prime\prime}}{hB^2}+\frac{h^{\prime}B^{\prime}}{hB^3}-\frac{h^{\prime}A^{\prime}}{hAB^2}-\frac{A^{\prime\prime}}{AB^2}+\frac{A^{\prime}B^{\prime}}{AB^3}.$$

\subsubsection{The torsion quantities}

Now we come to the torsional part and calculate the teleparallel connection
$$\Gamma^{\alpha}_{\mu\nu}=e_a^{\alpha}\partial_{\mu}e^a_{\nu}+e_a^{\alpha}\eta^{ac}\omega_{c\mu b} e^b_{\nu}.$$
A little thought shows that it is a very simple calculation, with the first and the second terms giving contribution to different components. In the pure tetrad approach, the following components would be non-zero:
$$\Gamma^0_{10}=\frac{A^{\prime}}{A}, \qquad \Gamma^1_{11}=\frac{B^{\prime}}{B}, \qquad \Gamma^2_{12}=\Gamma^3_{13}=\frac{h^{\prime}}{h}, \qquad \Gamma^3_{23}=\cot\theta,$$
and the spin connection (\ref{omega}) adds to that
$$\Gamma^2_{21}=-\Gamma^1_{22}=\frac{B}{h}, \qquad \Gamma^3_{31}=-\Gamma^1_{33}=\frac{B}{h}, \qquad 
\Gamma^3_{32}=\cot\theta, \qquad \Gamma^2_{33}=-\sin(\theta)\cos(\theta).$$

Then the torsion tensor has the following non-vanishing components:
$$T^{0}_{\hphantom{0}10}=-T^{0}_{\hphantom{0}01}=\frac{A^{\prime}}{A}, \qquad T^{2}_{\hphantom{0}12}=T^{3}_{\hphantom{0}13}=-T^{2}_{\hphantom{0}21}=-T^{3}_{\hphantom{0}31}=\frac{h^{\prime}}{h}-\frac{B}{h}.$$
The spin connection made the subtraction of $\frac{B}{h}$, and what is even more important, it made the $T^{3}_{\hphantom{0}23}$ component vanish.

We can now easily find the non-zero components of the contortion tensor (\ref{cont}) as
$$K_{100}=-K_{001}=AA^{\prime}, \qquad K_{221}=-K_{122}=h(h^{\prime}-B), \qquad K_{331}=-K_{133}=h(h^{\prime}-B)\sin^2\theta.$$
A good consistency check is to compare with the previous subsubsection and see that indeed $$\Gamma^{\alpha}_{\mu\nu}-\mathop\Gamma\limits^{(0)}{\vphantom{\Gamma}}^{\alpha}_{\mu\nu}(g)=K^{\alpha}_{\hphantom{\alpha}\mu\nu}.$$

Finally we see that only one component of the torsion vector is non-trivial, $T_1=\frac{A^{\prime}}{A}+2\left(\frac{h^{\prime}}{h}-\frac{B}{h}\right)$, and the superpotential is given by
$$S_{001}=-S_{010}=2A^2\left(\frac{h^{\prime}}{h}-\frac{B}{h}\right),$$ $$ S_{212}=-S_{221}=h(h^{\prime}-B)+h^2\frac{A^{\prime}}{A}, \qquad S_{313}=-S_{331}=\left(h(h^{\prime}-B)+h^2\frac{A^{\prime}}{A}\right)\sin^2\theta,$$
with other components vanishing. Of course, the only non-zero derivative of $\mathbb T$ is with respect to the radial coordinate $r$. Then note that $S_{\mu\nu 1}$ has only diagonal entries, and therefore the antisymmetric part of equations of motion (\ref{eom}) is automatically satisfied.

Note also that it is very easy to see why the diagonal tetrad (\ref{dtet}) without the spin connection doesn't work well. Indeed, we have then $T^{3}_{\hphantom{0}23}=\cot\theta\neq 0$ which adds another non-zero component to the torsion vector, $T_2=\cot\theta$, and therefore it produces a  problematic non-zero component $S_{121}$ of the superpotential (because it has the term $g_{11}T_2$ in it). As a result, the mixed $r$-$\theta$ component, but not the $\theta$-$r$ one, of equations of motion is not zero and requires either constant $\mathbb T$ or TEGR.

\subsubsection{The equations}

Now we have all we need. From now on, let's put $h(r)=r$, and therefore $h^{\prime}=1$ and $h^{\prime\prime}=0$, to compare with equations in the existing literature \cite{spher3, GoGu3}.

A very simple calculation gives then the torsion scalar:
\begin{equation*}
{\mathbb T}=\frac 12 S_{\alpha\mu\nu}T^{\alpha\mu\nu} = -2(B-1)\frac{2rA^{\prime}-A(B-1)}{r^2 A B^2}.
\end{equation*}
And let's write the equations (\ref{eom}) as
$$\frac12 f\delta^{\mu}_{\nu}+ f_{T}\left(\mathop{G^{\mu}_{\nu}}\limits^{(0)}-\frac12 {\mathbb T}\delta^{\mu}_{\nu}\right)+f_{TT}S^{\mu\hphantom{\mu}\alpha}_{\hphantom{\mu}\nu}\partial_{\alpha}{\mathbb T}=0.$$
Using the results from the previous two subsubsections, it gives the equations we had before \cite{spher3, GoGu3}: the temporal
\begin{equation}
\label{timeq}
\frac{f}{2} + \frac{2f_{T}}{r A B^3}\left(r(B-1)BA^{\prime}+rAB^{\prime} +AB(B-1))\vphantom{\int}\right) + \frac{2(B-1)}{r B^2} f_{TT}{\mathbb T}^{\prime}=0,
\end{equation}
the radial
\begin{equation}
\label{radeq}
\frac{f}{2} + \frac{2f_{T}}{r^2 A B^2} \left(r(B-2)A^{\prime}+A(B-1)\vphantom{\int} \right)=0,
\end{equation}
and the angular
\begin{equation}
\label{angeq}
\frac{f}{2} - \frac{f_{T}}{r^2 A B^3} \left(r^2(BA^{\prime\prime}-A^{\prime}B^{\prime})-r(2B-3)BA^{\prime}-rAB^{\prime}+AB(B-1)^2\vphantom{\int}\right) -\frac{rA^{\prime}-A(B-1)}{rA B^2}f_{TT}{\mathbb T}^{\prime}=0,
\end{equation}
now obtained by simple calculations on a few sheets of paper.

\section{Using the diffeomorphism invariance}

We have shown how the derivation of spherically symmetric equations in $f(T)$ can be made easy. Now let us come back to the questions of basic symmetries. As we have mentioned above, the diffeomorphism invariance is not broken in any version of the model. And it can be used with success. Here we would like to give two examples of this.

\subsection{Understanding why it is solvable at all}

One of immediate puzzles for anyone who looks at this topic is why the three equations can be solvable if we have only two unknown functions, $A(r)$ and $B(r)$. (The third one, the $h(r)$, cannot be of any help since it reflects just a mere change of coordinates.) Similar questions arise also in general relativity, and the answer is in the Bianchi identity which implies $\mathop{\bigtriangledown_{\mu}}\limits^{(0)}G^{\mu}_{\nu}=0$.

In modified teleparallel gravity the situation is different. However, using the diffeomorphism invariance in the action, it has been shown \cite{GoGu3} that, once the antisymmetric part of equations is satisfied, the rest of it also has vanishing divergence. It means that only two equations are independent.

Since only two equations are independent, we can also choose which ones to use. The radial one (\ref{radeq}) is of lower differential order, and therefore has to be used. Then, out of the remaining two, temporal (\ref{timeq}) and angular (\ref{angeq}), we can choose any. Even a better idea is to take a linear combination of them in which the $f_{TT}\mathbb T^{\prime}$ term cancels out. It gives an equation with $f$ and $f_T$ terms only. Viewed together with the radial equation (\ref{radeq}) as a system of two linear equations for $f$ and $f_T$, its matrix must be degenerate in order to have a non-zero solution, and therefore to not switch the gravity off. It can be used to find the following equation independent of the function $f$:
\begin{equation}
\label{findeq}
-r^2(B-1)AA^{\prime\prime}+ r^2 {A^{\prime}}^{2} + r^2 AA^{\prime} B^{\prime}-A^2(B+1)(B-1)^2=0.
\end{equation}

In our papers \cite{GoGu3, spher3} we derived this equation (\ref{findeq}) more directly by solving the radial equation (\ref{radeq}) for $f$ in terms of $f_T$ and substituting into the combination of Eqs. (\ref{timeq}) and (\ref{angeq}) which contains $f$ and $f_T$ only. Then one gets an equation with a common factor of $f_T$. And as long as $f_T\neq 0$, so that the gravity does not get switched-off, the factor can be cancelled bringing us to the result above (\ref{findeq}).

This is a very interesting observation. There is this relation (\ref{findeq}) between the functions $A(r)$ and $B(r)$ which must be satisfied for the tetrad (\ref{sphtet}) to be a solution of $f(T)$ gravity with practically every function $f$. The only exceptions are the cases with constant $\mathbb T$ fixed to the point of $f_T=0$ where the gravity is, in a sense, removed at all. Some examples of this pathology are given in the Ref. \cite{GoGu3}.

In presence of matter we would have to put the energy-momentum tensor to the right hand side of the Eq. (\ref{eom}). It would not allow us to get the equation (\ref{findeq}) purely independent of $f$. Instead, its right hand side would be not zero but a linear combination of density and pressure (with coefficients depending on $A$ and $B$ though) divided by $f_T$. This is absolutely not surprising, since the strength of coupling to matter must play its role for sure.

\subsection{Diagonal tetrad with zero spin connection is possible}

Also, diffeomorphism invariance can help us understand that diagonal tetrads in pure tetrad approach are still possible. Indeed, the tetrad (\ref{sphtet}) works well with any choice of the radial coordinate without a spin connection, and in particular we can use the $B=h$ case. Having done so, we see that the new components 
\begin{equation}
\label{Bsphertet}
e^{a}_{\mu} = \left(
\begin{array}{cccc}
A(r) & 0 & 0 & 0\\
0 & B(r) \sin(\theta)\cos(\phi) & B(r) \cos(\theta) \cos(\phi) & -B(r)\sin(\theta) \sin(\phi) \\
0 & B(r) \sin(\theta) \sin(\phi) & B(r) \cos(\theta) \sin(\phi) & B(r) \sin(\theta) \cos(\phi) \\
0 & B(r) \cos(\theta) & -B(r)\sin(\theta) & 0
\end{array}
\right)
\end{equation}
of the tetrad (\ref{sphtet}) can be seen as components of a trivial Cartesian tetrad written in spherical coordinates. It means that this tetrad will become diagonal by the change of coordinates to the Cartesian ones.

And indeed, as it has been noticed in the Ref. \cite{GoGu3}, if we take the metric as 
\begin{equation}
\label{Carmet}
g_{\mu\nu}dx^{\mu}dx^{\nu}=A^2(r)dt^2-B^2(r)(dx^2+dy^2+dz^2)
\end{equation}
with $r\equiv \sqrt{x^2+y^2+z^2}$ and the diagonal tetrad 
\begin{equation}
\label{Cartet}
e^a_{\mu}=\mathrm{diag}(A,B,B,B),
\end{equation}
then it also solves the antisymmetric part of equations. And, of course, the equations would be equivalent to the ones we had above, modulo the change of the radial variable.

Let us illustrate also here that it won't be difficult to see that the equations are solvable, or that the tetrad is "good" in the common jargon. Indeed, the calculations of the non-symmetric term go even simpler. The connection is
$$\Gamma^0_{i0}=\frac{A^{\prime}x_i}{rA}, \qquad \Gamma^k_{ij}=\frac{B^{\prime}x_i}{rB}\delta^k_j$$
which gives the non-zero torsion tensor components as
$$T_{0i0}=\frac{1}{r}AA^{\prime}x_i, \qquad T_{kij}=\frac{B^{\prime}}{rB}(g_{kj}x_i-g_{ki}x_j)$$
with the torsion vector $T_i=\frac{x_i}{r}\left(\frac{A^{\prime}}{A}+2\frac{B^{\prime}}{B}\right)$ and the contortion tensor
$$K_{i00}=-K_{00i}=\frac{1}{r}AA^{\prime}x_i, \qquad K_{ijk}=\frac{B^{\prime}}{rB}(g_{ij}x_k-g_{jk}x_i)$$
easily found.

Then the superpotential takes the form of
$$S_{00i}=-S_{0i0}=2\frac{A^2 B^{\prime}}{rB}x_i, \qquad S_{ijk}=\left(\frac{A^{\prime}}{rA}+\frac{B^{\prime}}{rB}\right)(g_{ij}x_k-g_{ik}x_j).$$
We see that the spatial part of the superpotential does have a non-symmetric (in the first two indices) part proportional to $g_{ik}x_j$. However, since the torsion scalar $\mathbb T$, as everything else, depends only on $r$, the $f_{TT}$ term in the equations of motion (\ref{eom}) given by $S_{ijk}\partial^k \mathbb T$ has a purely diagonal contribution and also a term proportional to $x_i x_j \mathbb T^{\prime}$ which turns out to be symmetric, and moreover, of the same $\propto x_i x_j$ shape as non-diagonal pieces in the Einstein tensor in these coordinates.

\section{Comparing the two formulations again}

Now, we are coming to the questions of Lorentz invariance. It was claimed in the Ref. \cite{SariCai} that the covariant version of $f(T)$ is better because it doesn't have the problem of non-uniqueness of the tetrad. It is wrong. It has precisely the same number of physically different solutions as the pure tetrad formulation does. In no sense is physical uniqueness of solution obtained by covariantisation.

If we found a solution in the covariantised version, then taking the $\omega=0$ gauge we get a solution in the pure tetrad formulation. If we found a solution in the latter, it is also a solution in the covariant version. Moreover, if a diagonal tetrad at all exists for this metric, we can find a corresponding spin connection, and different solutions of the pure tetrad model would then correspond to different possible spin connections with the same diagonal tetrad. The covariant $f(T)$ gravity does not define a unique spin connection for a given tetrad.

\subsection{Another way to spherically symmetric solutions}

The uniqueness claim of the Ref. \cite{SariCai} was based on the particular spin connection (\ref{omega}) which can be used with the diagonal tetrad (\ref{dtet}) and corresponds then to the tetrad (\ref{sphtet}) without a spin connection. However, it is not the only possible spin connection. Since the pure tetrad formulation has also another possible tetrad of the Ref. \cite{spher4}, we can also construct the spin connection which, taken with the diagonal tetrad (\ref{dtet}), would correspond to this tetrad.

Indeed, this alternative tetrad can be written as
\begin{equation}
\label{Isphertet}
e^{a}_{\mu} = \left(
\begin{array}{cccc}
A(r) & 0 & 0 & 0\\
0 & B(r) \sin(\theta)\cos(\phi) & - r \cos(\theta) \cos(\phi) & r\sin(\theta) \sin(\phi) \\
0 & B(r) \sin(\theta) \sin(\phi) & - r \cos(\theta) \sin(\phi) & - r \sin(\theta) \cos(\phi) \\
0 & B(r) \cos(\theta) & r\sin(\theta) & 0
\end{array}
\right),
\end{equation}
different from the tetrad (\ref{sphtet}) by the signs of the right two columns. It corresponds to the Lorentz rotation from the diagonal tetrad (\ref{dtet}) slightly different from the one (\ref{Lor}) we used before:
\begin{equation*}
\Lambda = \left(
\begin{array}{cccc}
1 & 0 & 0 & 0 \\
0 &  \sin(\theta)\cos(\phi) &  -\cos(\theta) \cos(\phi) &  \sin(\phi) \\
0 &  \sin(\theta) \sin(\phi) &  -\cos(\theta) \sin(\phi) &  -\cos(\phi) \\
0 &  \cos(\theta) & \sin(\theta) & 0
\end{array}
\right),
\end{equation*}
which can also be easily inverted as
\begin{equation*}
\Lambda^{-1} = \left(
\begin{array}{cccc}
1 & 0 & 0 & 0 \\
0 &  \sin(\theta)\cos(\phi) &  \sin(\theta) \sin(\phi) &  \cos(\theta) \\
0 &  - \cos(\theta) \cos(\phi) &  - \cos(\theta) \sin(\phi) &  \sin(\theta) \\
0 &  \sin(\phi) & - \cos(\phi) & 0
\end{array}
\right)
\end{equation*}
what gives
$$\Lambda^{-1}\partial_{\theta}\Lambda=\left(
\begin{array}{cccc}
0 & 0 & 0 & 0 \\
0 &  0 & 1 &  0 \\
0 &  -1 & 0 & 0 \\
0 & 0 & 0 & 0
\end{array}
\right), \qquad\Lambda^{-1}\partial_{\phi}\Lambda=\left(
\begin{array}{cccc}
0 & 0 & 0 & 0 \\
0 &  0 &  0 &  \sin\theta \\
0 &  0 & 0 & -\cos\theta \\
0 & -\sin\theta & \cos\theta & 0
\end{array}
\right)$$
and corresponds to $\omega^2_{\hphantom{2}\theta 1}=-1, \quad \omega^3_{\hphantom{2}\phi 1}=-\sin\theta, \quad \omega^3_{\hphantom{2}\phi 2}=\cos\theta$.

In other words, in the covariantised version of $f(T)$ gravity the diagonal tetrad (\ref{dtet}) can also be used with the following spin connection:
\begin{equation}
\label{minomega}
\omega_{1 \theta 2}=-\omega_{2 \theta 1}=-1,\qquad \omega_{1 \phi 3}=-\omega_{3 \phi 1}=-\sin\theta,\qquad \omega_{2 \phi 3}=-\omega_{3 \phi 2}=\cos\theta.
\end{equation}
This is another possible choice of the spin connection for the diagonal tetrad (\ref{dtet}), and it actually corresponds to the pure tetrad of the Ref. \cite{spher4} and to another pure tetrad case of the Ref. \cite{SariCai}, once again showing the equivalence of the two formulations of the theory.

And, actually, in this case it is very simple to see that there is indeed such an option by simply looking at our derivations from above. From the teleparallel connection coefficients we have found in the subsubsection 3.1.2, it is obvious that for vanishing the problematic $T^{3}_{\hphantom{0}23}$ torsion component we need to keep $\omega_{2\phi 3}$ the same. However, changing the signs of $\omega_{1\theta 32}$ and $\omega_{1\phi 3}$ would result only in change of the sign in front of $B$ which cannot spoil the satisfaction of the antisymmetric equations, and it is very simple to check that the new spin connection is also flat as it must be. Modulo unimportant overall sign of the tetrad, this change of $B$ to $-B$ is precisely how the tetrad (\ref{Isphertet}) differs from our previous tetrad (\ref{sphtet}); note that the sign of $A$ plays no role, for it is just a mere reversal of time.

So, we have shown that the covariant formulation also has various possible solutions, the same way as the pure tetrad formulation does. In the covariant formulation, it can be viewed in different ways, either like in the pure tetrad one by looking at the possible tetrads with zero spin connection (or with another fixed spin connection), or by choosing some tetrad at will and looking at which spin connections are compatible with this choice, or by using any reasonable mixture of those two ways.

The actual cause of the opposite opinion expressed in the Ref. \cite{SariCai} is that they take only one of many possible solutions for $\omega$. Their choice is in terms of a reference tetrad with switched-off gravity \cite{spinc, SariCai}. But there is nothing in the equations of motion which would force us to make this very choice. We saw that not only the spin connection (\ref{omega}), but also another one (\ref{minomega}) is compatible with the diagonal tetrad (\ref{dtet}). And of course, like in the Ref. \cite{SariCai}, we can use the tetrad (\ref{sphtet}) with zero spin connection and the tetrad (\ref{Isphertet}) with a spin connection which corresponds to the local Lorentz transformation from one to another; but we can also use zero spin connection with the tetrad (\ref{Isphertet}) and a spin connection obtained from the reverse transformation between these two tetrads for the tetrad (\ref{sphtet}); and those two options are again the same as our two choices of the spin connection for the diagonal tetrad (\ref{dtet}).

There is nothing in the equations of motion which would require one particular choice of the spin connection related to switching off gravity. And it was actually noticed \cite{MT} that there is even an ambiguity in this recipe, and in cases beyond the static spherical symmetry it can even give incorrect result which does not satisfy the antisymmetric part of equations of motion \cite{add4}.

Of course if, due to any reason, one starts worrying about the value of the action, not only about the equations of motion, then there can be certain physical grounds \cite{spinc} for choices in terms of a reference tetrad. But this is also true of the pure tetrad approach. If we take the Minkowski limit of $A=B=1$ with another tetrad (\ref{Isphertet}), or equivalently we go for $A=B=-1$ in the standard tetrad (\ref{sphtet}), then we get the Minkowski metric but with non-zero $\mathbb T$ and not even a solution. In this sense, it is enough a reason already in the pure tetrad formulation to prefer the tetrad (\ref{sphtet}), or to go for negative functions $A$ and $B$ in the tetrad (\ref{Isphertet}) which would be the same.

\subsection{Symmetric solutions with constant $\mathbb T$ and/or without symmetry}

Finally, let us note that, at the level of the tetrad ansatz, both cases above can be treated at once using the formula (\ref{sphtet}) if we don't fix some particular sign of the function $B$. Though, due to non-linearity, there might be several different solutions, even with a particular choice of the sign of $B$. However, why it is indeed, in a certain sense, almost unique is because we were looking for a really symmetric solution, not only at the level of having symmetry in the metric.

Spherically symmetric solutions were searched for in $f(T)$ gravity from its very beginning \cite{FerFio}, and for a long time these searches were mostly going in the direction of making the torsion scalar vanish, or at least be constant \cite{Gamal}. Once the torsion scalar is constant, the solution of equations (\ref{eom}) is obviously equivalent to GR with a cosmological constant (not at the level of perturbations though).
The paper \cite{FerFio} is very interesting in this respect. It is one of the first ones on this topic, and the Authors are looking for the Schwarzschild solution. For functions $f$ beyond TEGR, it requires constant $\mathbb T$ of course. However, a strange feature is that they switch to Cartesian coordinates for having the antisymmetric part of equations solved with the diagonal tetrad (\ref{Cartet}) and without a spin connection. After that they make a Lorentz boost which brings $\mathbb T$ to zero value, and therefore allows for Schwarzschild geometry as a solution. 

Of course, going for constant $\mathbb T$ could have been performed also in the spherical coordinates with a local Lorentz transformation of the diagonal tetrad (\ref{dtet}) because the antisymmetric part of equations always disappears when the torsion scalar is made to be constant. However, it won't be so easy to achieve that starting from the tetrad (\ref{dtet}) if to stick to the spherical symmetry at the level of the tetrad. The Lorentz transformation $
\Lambda = \left(
\begin{array}{cccc}
\cosh\lambda(r) & \sinh\lambda(r) & 0 & 0 \\
\sinh\lambda(r) &  \cosh\lambda(r) &  0 & 0 \\
0 &  0 &  1 &  0 \\
0 & 0 & 0 & 1
\end{array}
\right)$ applied to the diagonal tetrad (\ref{dtet}) does not change the torsion scalar at all, and therefore can be called a remnant symmetry \cite{remn}. It can be seen by simply calculating the corresponding spin connection instead of really going for a non-diagonal tetrad. Obviously, the only non-zero component is $\omega_{0r1}=-\omega_{1r0}$ which only adds a non-zero time component to the torsion vector, $T_0(r)$, and therefore, being a time component depending only on the radius with a diagonal metric, it does not change $\mathop{\bigtriangledown_{\mu}}\limits^{(0)}T^{\mu}$, and due to the relation (\ref{relation}) cannot change $\mathbb T$ either.

One can always play with violating the symmetry at the level of the tetrad, and a rather simple idea would be to arrange for constant $\mathbb T$ indeed. The covariant version of $f(T)$ does not make us free of this crazy zoo of solutions with spherically symmetric metrics. And, by the same procedure as above, we can find a complicated and non-symmetric spin connection which, being used with the diagonal tetrad (\ref{dtet}), would give us the same solution as we were looking at. 

All in all, one can take any of these constant $\mathbb T$ solutions for the pure tetrad $f(T)$ spherically symmetric vacuum geometries. After rewriting in spherical coordinates, if needed, there would be a Lorentz transformation $\Lambda$ to bring the simplest tetrad (\ref{dtet}) to that one. And in covariantised $f(T)$ any such solution can be described as this tetrad (\ref{dtet}) with the spin connection $\omega=\Lambda^{-1}\partial\Lambda$. 

If we require the symmetry at the level of the metric only, then there are infinitely many solutions, and many of them with constant $\mathbb T$. In particular, any spin connection which makes $\mathbb T$ constant for a chosen tetrad would solve the antisymmetric equations, and this requirement is just about one scalar quantity to vanish when having six free parameters of the Lorentz transformations for achieving this goal.

\section{Conclusions}

In the course of discussing the Lorentz-(non-)invariance, we have explicitly shown how to easily make computations for studying the spherically symmetric solutions in $f(T)$ gravity, using the benefits of the covariantised approach, despite it being fully equivalent to the pure tetrad model. 

A simple and direct way would be to look at the spherically symmetric tetrad as a set of quite trivial vectors in Cartesian coordinates (\ref{Cartet}), in analogy with an old idea of vector inflation \cite{meold}. This is already good enough, and one can see that antisymmetric equations are satisfied with zero spin connection. It can be solved this way.

But a natural wish is to go for spherical coordinates. The change of coordinates from Cartesian to the spherical ones transforms the tetrad (\ref{Cartet}) into the tetrad (\ref{Bsphertet}) which can be turned into the standard one (\ref{sphtet}) by a change of the radial coordinate. 

At the same time, one can easily check that the simple diagonal tetrad in spherical coordinates (\ref{dtet}) is not good enough in the pure tetrad approach. But instead of a complicated calculation with the non-diagonal tetrad (\ref{sphtet}), we can find the flat spin connection (\ref{omega}) which, being used with the diagonal tetrad (\ref{dtet}), is equivalent to the pure tetrad (\ref{sphtet}). And then the calculations are much simpler.

Moreover, while doing our calculations, it was easy to see that a spin connection (\ref{minomega}) can also be used with the diagonal tetrad (\ref{dtet}), bringing the case of another pure tetrad (\ref{Isphertet}) into the covariant approach, too. This is an explicit demonstration of how the two approaches to modified teleparallel gravity are equivalent to each other.

{\bf Acknowledgements.} I am grateful to Mar{\'i}a Jos{\'e} Guzm{\'a}n and Adel Awad for numerous discussions on the topics of modified teleparallel gravity.

\end{document}